\documentclass[journal,transmi]{IEEEtran}

\usepackage{cite}

\usepackage{graphicx}
\usepackage{float}
\usepackage{amsmath}
\usepackage{amssymb}
\newtheorem{theorem}{Theorem}
\usepackage{algorithm}
\usepackage{algorithmic}
\usepackage{array}
\ifCLASSOPTIONcompsoc
  \usepackage[caption=false,font=normalsize,labelfont=sf,textfont=sf]{subfig}
\else
  \usepackage[caption=false,font=footnotesize]{subfig}
\fi

\usepackage{stfloats}

\usepackage{url}

\usepackage{multirow}
\newcommand{\minitab}[2][l]{\begin{tabular}{#1}#2\end{tabular}}
\usepackage{color, colortbl}

\begin{document}

\title{Detecting abnormal connectivity in schizophrenia via a joint directed acyclic graph estimation model}

\author{\IEEEauthorblockN{Gemeng Zhang\IEEEauthorrefmark{1},
Aiying Zhang\IEEEauthorrefmark{1},
Biao Cai\IEEEauthorrefmark{1},
Zhuozhuo Tu\IEEEauthorrefmark{2},
Vince D. Calhoun\IEEEauthorrefmark{3} 
and Yu-Ping Wang\IEEEauthorrefmark{1}}\\
\IEEEauthorblockA{\IEEEauthorrefmark{1}Department of Biomedical Engineering, Tulane University, New Orleans, LA 70118 USA}\\
\IEEEauthorblockA{\IEEEauthorrefmark{2}UBTECH Sydney Artificial Intelligence Centre, The University of Sydney, NSW, 2006 Australia}\\
\IEEEauthorblockA{\IEEEauthorrefmark{3}Tri-institutional Center for Translational Research in Neuroimaging and Data Science (TReNDS) Georgia State University, Georgia Institute of Technology, Emory University, Atlanta, GA 30030 USA}}

\IEEEtitleabstractindextext{
\begin{abstract}
 Functional connectivity (FC) has been widely used to study brain network interactions underlying the emerging cognition and behavior of an individual. FC is usually defined as the correlation or partial correlation between brain regions. Although FC is proved to be a good starting point to understand the brain organization, it fails to tell the causal relationship or the direction of interactions. Many directed acyclic graph (DAG) based methods were applied to study the directed interactions using functional magnetic resonance imaging (fMRI) data but the performance was severely limited by the small sample size and high dimensionality, hindering its applications. To overcome the obstacles, we propose a score based joint directed acyclic graph model to estimate the directed FC in fMRI data. Instead of using a combinatorial optimization framework, the structure of DAG is characterized with an algebra equation and further regularized with sparsity and group similarity terms. The simulation results have demonstrated the improved accuracy of the proposed model in detecting causality as compared to other existing methods. In our case-control study of the MIND Clinical Imaging Consortium (MCIC) data, we have successfully identified decreased functional integration, disrupted hub structures and characteristic edges (CtEs) in schizophrenia (SZ) patients. Further comparison between the results from directed FC and undirected FC illustrated the their different emphasis on selected features. We speculate that combining the features from undirected graphical model and directed graphical model might be a promising way to do FC analysis. The code of joint DAG estimation model is available at \url{https://github.com/gmeng92/joint-notears}.
\end{abstract}
\begin{IEEEkeywords}
task fMRI, causality, joint estimation, schizophrenia
\end{IEEEkeywords}}
\maketitle

\IEEEdisplaynontitleabstractindextext

\section{INTRODUCTION}\label{sec:intro}  
For the past decades, the research of brain functional connectivity (FC) through brain imaging data has became one of the hot topics in medical imaging field\cite{jafri2008method}. Specifically, studies using functional MRI (fMRI) have shown that there are significant connectivity changes in mental disorder patients compared to normal cohort\cite{calhoun2009functional}\cite{cerliani2015increased}. As a serious mental illness, schizophrenia has also been conceived as a disorder with FC changes in large-scale brain networks\cite{lynall2010functional}. Previous meta-analytic reviews of MRI studies on schizophrenia also suggest the underlying abnormalities in gray matter density\cite{glahn2008meta}\cite{ellison2008anatomy}, as well as the interregional FC derived from fMRI time series\cite{liang2006widespread}\cite{salvador2010overall}\cite{ma2019decreased}.

As a popular tool for accessing functional organization of brain and as an important biomarker for neurological disorders, functional connectivity is often defined based on the Pearson correlation which reflects the association between different regions of interests (ROIs). Although further studies used the partial correlation or precision matrix to remove the indirect correlation between brain regions and only keep the directly correlated pairs, they still cannot reveal the causes and effects between brain regions. More precise understanding of the information flow between brain regions enables us to better understand brain functional integration. The greedy equivalent search (GES) method\cite{meek1997graphical}\cite{chickering2002optimal}, as an efficient way to do causal inference, has been applied on fMRI data to investigate differences in brain integrations between patients with Autism spectrum disorders\cite{hanson2013atypical} and individuals with traumatic brain injury\cite{dobryakova2015investigation}. By modeling the non-Gaussianity of the real fMRI data, researchers have also employed the linear non-gaussian acyclic models (LiNGAM)\cite{shimizu2006linear} to reveal the key differences in the default-mode network between patients with bipolar disorder and patients with major depression disorder\cite{liu2015altered}.

Recently, there is a score-based approach for the structure learning of a directed acyclic graph (DAG)\cite{zheng2018dags}. Different from other greedy equivalence search approaches, the new method incorporates the DAG constraints by introducing an algebraic characterization of the adjacency matrix of one DAG, which gives more flexibility in modeling the structure of Bayesian networks. For example, the authors enforced the sparsity of the DAG structure and got the state-of-the-art result in their simulation data. This motivates us to apply the similar idea for fMRI data analysis.

However, the high dimensionality and moderate individual variations prevent the score-based method from fitting well on the fMRI data since most score criteria are defined as the likelihood of the Bayesian networks. To mitigate the limitation when applying to the fMRI analysis, we further formulate a joint structure learning framework, in which each DAG's structure estimation is performed by borrowing information from other observations or groups. Although the proposed structure learning model is not a convex one, we found that using the augmented Lagrange and L-BFGS method can deliver significantly improved results. The performance of the model is validated through a series of simulations including various settings for random graphs and compared with four other widely-used DAG estimation methods.

 The rest of the paper is organized as follows. In Section \ref{sec:2}, we give some preliminary knowledge of Bayesian networks, formulate joint DAGs estimation model and describe the optimization strategy. In Section \ref{sec:3}, we show the results from both simulation studies and fMRI studies.  Then we further present regarding the findings and limitations of the proposed method in Section\ref{sec:4} and conclude the paper is in the last section.

\section{METHODS}\label{sec:2}
\subsection{Preliminary}
A Bayesian network is a probabilistic graphical model that encodes the random variables as its nodes and their conditional dependencies as its directed edges. The node set $V$ and the edge set $E = V\times V$ make up with a directed acyclic graph (DAG) $G = (E,V)$ in which each edge $(i,j)$ represents a directed edge from variable $i$ to $j$. $i$ is called the parent node of $j$ and $j$ is the child node of $i$ accordingly. Given one variable, say $j$, and the set of all of its parent nodes $Pa_{j}$, we can always write the conditional probability distribution of $j$ as $P(j|Pa_{j})$.Without loss of generality, we can always factorize the joint distribution of random variables $Y= \{y_1, y_2, \dots, y_d\}$ as $P(y) = \prod_{i=1}^{d}P(y_i|Pa_{i})$.

Given the data matrix $X\in\mathbb{R}^{n\times d}$ as the $n$ observations of random variables $y\in \mathbb{R}^d$, learning the structure of the Bayesian network or directed acyclic graph (DAG) means finding a proper distribution $P(X)$ defined on the graph $G = (V,E)$. More specifically, we want to determine all the conditional dependencies via the observation $X$. We model these conditional dependencies via a structural equation model(SEM) defined by $X= XW+z$, where $W\in\mathbf{R}^{d\times d}$ is the weighted adjacency matrix and $z$ is the random noise vector. Without further assumption on the noise distribution, we follow the approach in \cite{zheng2018dags} and use the least square loss function for the linear SEM. It should be noted that the loss can be changed to any other smooth functions over $\mathbf{R}^{d\times d}$. Moreover, several previous studies have shown that minimizing the least square loss can guarantee revealing the true DAG with high probability in finite-samples and even in high dimensions setting($d\gg n$) and this result is consistent for both Gaussian SEM\cite{aragam2015learning}, \cite{van2013ell_} and non-Gaussian SEM\cite{loh2014high}.

In the literature\cite{zheng2018dags}, the authors firstly characterized the adjacency matrix of a DAG algebraically and turned the fussy searching problem in traditional GES methods into a continuous optimization issue. Their characterization is formulated with the theorem
\begin{theorem}\cite{zheng2018dags}
  A matrix $W\in\mathbb{R}^{d\times d}$ is an adjacency matrix of a DAG if and only if
  \begin{equation*}
    h(W) = Tr(e^{W\circ W})-d = 0,
  \end{equation*}
\end{theorem}
where $\circ$ is the Hadamard product and $e^{\cdot}$ is the matrix exponential operator. Fortunately, this function is differentiable with a simple gradient
\begin{equation*}
  \nabla h(W) = 2(e^{W\circ W})^T\circ W,
\end{equation*}
\subsection{Optimization}
To overcome the sample size limitation and take advantage of data from multiple groups, we propose a joint DAG estimation model to incorporate the similarity prior in the data. The primal problem can be formulated as
\begin{equation}\label{eqn: joint_ori}
  \begin{aligned}
    \min_{\mathbf{W}} & &F(\mathbf{W}) = l(\mathbf{W};\mathbf{X}) + P(\mathbf{W})\\
    \text{s.t.} & &h(W^{(k)}) = 0,\ \ \forall k \in [K]
  \end{aligned}
\end{equation}
where $ l(\mathbf{W};\mathbf{X})$ is the loss function of SEM, i.e. $l(\mathbf{W};\mathbf{X}) = \sum_{k = 1}^{K}\frac{1}{2n_k}||X^{(k)}- X^{(k)}W^{(k)}||_{F}^{2}$. $P(\mathbf{W})$ is the penalty term of the weighted adjacency matrices and usually chosen to encourage $W^{(1)},W^{(2)},\dots,W^{(K)}$ to share certain characteristics such as the similar pattern of nonzero elements. In addition, the sparsity of those matrices are usually encoded as a prior knowledge which benefits both model training and interpretation of the results. Considering the similar causal structure underlining different observations, we borrow the idea of group regularization from the undirected graphical model to encourage the shared DAG structures. This technique has been proven useful in many gene expression network studies. The group regularization term is often formulated as:
\begin{equation}\label{eqn: group}
  P(\mathbf{W}) = \lambda_1\sum_{k=1}^{K}\sum_{i\neq j}|W^{(k)}_{ij}| + \lambda_2\sum_{i\neq j}(\sum_{k=1}^{K}(W^{(k)}_{ij})^2)^{1/2}
\end{equation}
where $\lambda_1$ and $\lambda_2$ are nonnegative parameters. $\lambda_1$ controls the sparsity of $W^{(k)}$s and the larger $\lambda_1$ is, the sparser the solution of (\ref{eqn: joint_ori}) will be. On the other side, $\lambda_2$ restrains the patterns of the nonzeros in $W^{(k)}$s and the larger $\lambda_2$ is, the more identical $W^{(k)}$s will be. If $\lambda_1$ and $\lambda_2$ are set to $0$, the model is degraded to the original notears model.

By applying the augmented Lagrange, the primal problem is written as
\begin{equation}\label{eqn:auglag}
\begin{aligned}
  \min_{\mathbf{W}}&  &l(\mathbf{W};\mathbf{X}) +P(\mathbf{W}) + \frac{\rho}{2}\sum_{k=1}^{K}|h(W^{(k)})|^2\\
  \text{s.t.}&  &h(w^{(k)}) = 0,\ \ \forall k \in [K]
  \end{aligned}
\end{equation}
We use the dual ascent to solve problem (\ref{eqn:auglag}) and the lagrange multiplier is derived as
\begin{equation}\label{eqn:lag_multi}
\begin{aligned}
  L(\mathbf{W},\alpha_1,\alpha_2,\dots,\alpha_K) = &l(\mathbf{W};\mathbf{X}) +P(\mathbf{W}) +\frac{\rho}{2}\sum_{k=1}^{K}|h(W^{(k)})|^2\\ &+\sum_{k=1}^{K}\alpha_kh(W^{(k)})
  \end{aligned}
\end{equation}
Hence, the corresponding dual function is
\begin{equation}\label{eqn:lag_dual}
  g(\alpha_1,\alpha_2,\dots,\alpha_K) = \min_{\mathbf{W}}L(\mathbf{W},\alpha_1,\alpha_2,\dots,\alpha_K)
\end{equation}
The dual problem is then
\begin{equation}\label{eqn:dual}
\max_{\alpha_1,\alpha_2,\dots,\alpha_K}g(\alpha_1,\alpha_2,\dots,\alpha_K)
\end{equation}
Denote $\mathbf{W}^{*}_{\mathbf{\alpha}}$ as the local minimizer of problem (\ref{eqn:lag_dual}), i.e. $g(\alpha_1,\alpha_2,\dots,\alpha_K) = L(\mathbf{W}^{*}_{\mathbf{\alpha}},\alpha_1,\alpha_2,\dots,\alpha_K)$. By noting that $g(\alpha_1,\alpha_2,\dots,\alpha_K)$ is a linear function of $\alpha_k$s, the partial derivatives are given by $\partial g(\alpha_1,\alpha_2,\dots,\alpha_K)/ \partial \alpha_k = h(W^{(k)^{*}})$. Hence we can perform the dual ascent by updating $\alpha_k$s with
\begin{equation}\label{eqn:update_alpha}
  \alpha_{k} \rightarrow \alpha_{k} + \rho h(W^{(k)^{*}})
\end{equation}
The rest of the concern will be the unconstrained optimization of subproblem (\ref{eqn:lag_dual}). Due to its high dimensionality and non-convexity, we follow the similar idea in \cite{zheng2018dags} and solve it with L-BFGS algorithm\cite{byrd1995limited} when $\lambda_1=\lambda_2 = 0$. When $\lambda_1, \lambda_2 >0$, the problem can be solved with proximal quasi-Newton (PQN) method\cite{zhong2014proximal}. As an iteration algorithm, at the $k$-th step, the descent direction is searched through a quadratic approximation of the smooth term:
\begin{equation}
  \label{eqn: direction}
  \mathbf{d}_k = \arg\min_{\mathbf{d}\in \mathbb{R}^p} \mathbf{g}_k^{T}\mathbf{d} + \frac{1}{2}\mathbf{d}^{T}B_k\mathbf{d} + \lambda_1||\mathbf{w}_k + \mathbf{d}||_1
\end{equation}
where $\mathbf{g}_k$ is the gradient of $f(w)$ and $B_k$ is the L-BFGS approximation of the Hessian matrix of $f(w)$. In addition, for each coordinate $j$, the solution of problem (\ref{eqn: direction}) has a closed form update $\mathbf{d} \leftarrow \mathbf{d} + z^* e_j$ in which $e_j$ is the unit vector in standard basis and
\begin{equation}
  \label{eqn: update}
  \begin{aligned}
  z^* &= \arg\min_{z} \frac{1}{2} \underbrace{B_jj}_{a} z^2 + \underbrace{(\mathbf{g}_j + (B\mathbf{d})_j)}_b z + \lambda_1|\underbrace{\mathbf{w}_j + \mathbf{d}_j}_c + z| \\
  & = -c + S_{c-b/a}(\lambda_1/a)
  \end{aligned}
\end{equation}
It should be noted that the low-rank structure of $B_k$ makes it sparse and fast to compute during the coordinate update. The sparse regularization in the model can further enable us to speed up the computation. Instead of updating for all coordinates, an active set of coordinates can be chosen based on the subgradients and we just need to update regarding the active set. More details of the subproblem optimization can be find in Appendix A.

\subsection{Parameter Tuning}
As it shows in the joint estimation model, there are two parameters,  i.e., $\lambda_1$ and $\lambda_2$  that need to be specified before we use it for causality inference. Different setting of those parameters will lead to different causal structures. Training the model with a large $\lambda_1$ will lead to a sparse $\mathbf{W}$ but may underfit the data. On the other hand, the model with a smaller $\lambda_1$ gives rise to a denser $\mathbf{W}$ but may overfit the data. Consequently, improper choice of the parameter will lead to inaccurate inference result: a sparser causal graph tends to have more false negatives while a denser graph usually has severe false positives problem. Hence we would like to balance both the simplicity of the model and the goodness of fit. Similarly, the parameter $\lambda_2$ reflects how similar the DAG structures will be between groups.  A larger $\lambda_2$ implies more similar structures between groups and vice versa. When $\lambda_2$ is trivially set to $0$, the joint estimation model is then degraded to $K$ seperate NOTEARS models for $K$ groups; As $\lambda_2$ goes to $\infty$, group variations will be wiped out gradually and the resulted $K$ DAGs will be identical to each other which is exactly the case that we regard the data as one group and apply the NOTEARS model directly. To find a proper combination of $(\lambda_1, \lambda_2)$, we did a 5-fold cross-validation using the value of $l(\mathbf{W})$ as the performance measure.
\begin{figure*}[htbp]
  \centering
  \includegraphics[width=0.98\textwidth]{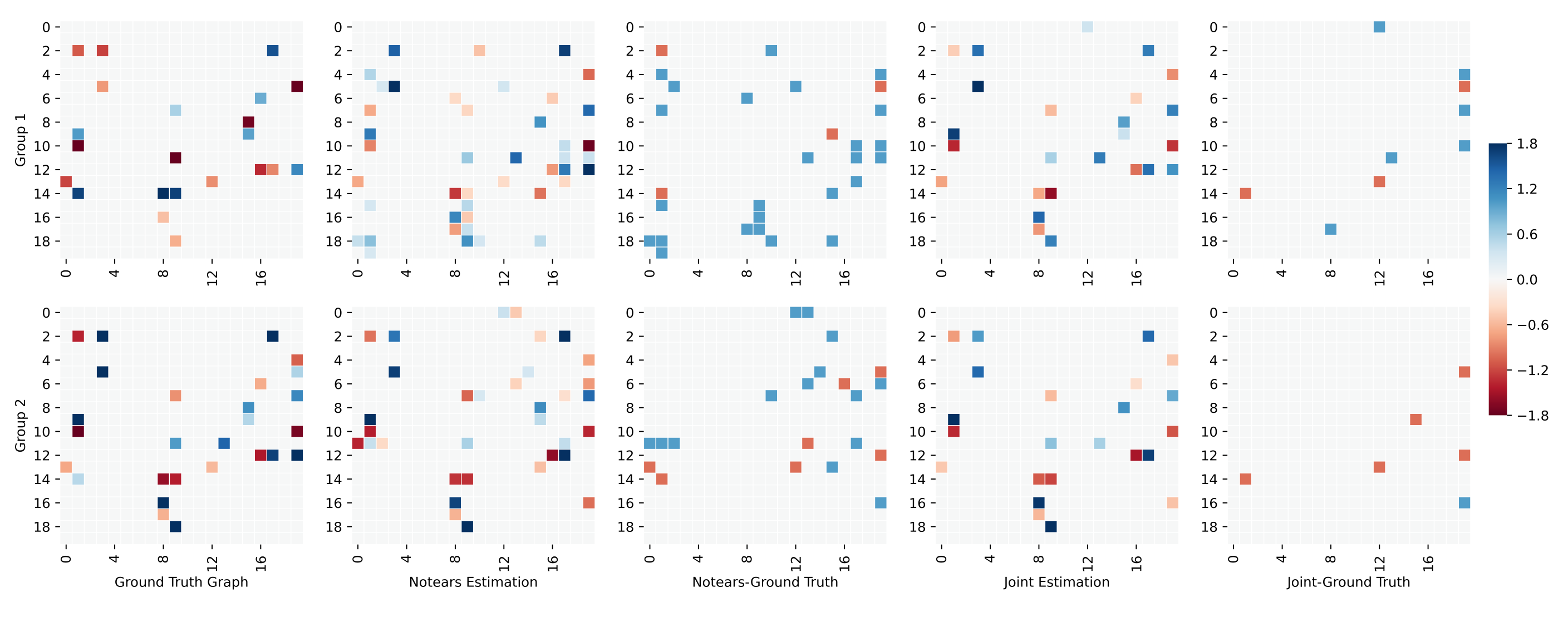}
  \caption{A demonstration of the joint estimation method compared to the NOTEARS method. The first row and second row show the weighted DAGs from group 1 and group 2, respectively. From left to right: the ground truth graphs, DAGs estimated by NOTEARS, the variation of NOTEARS estimation compared to ground truth, DAGs estimated by joint estimation, the variation of joint estimation compared to ground truth. In the 3rd and 5th column, the less the colored squares are, the better the performance the respective method has.}
  \label{fig:simu_demo}
\end{figure*}
\section{RESULTS}\label{sec:3}
\subsection{On Simulation Data}
\subsubsection{Data Generating}
To validate the performance of our joint DAGs estimation method, we evaluated it with the generated simulation data sets under different circumstances. We also compared our methods with the score-based greedy equivalent search (GES)\cite{chickering2002optimal}, the PC algorithm\cite{spirtes2000causation}, and the linear non-Gaussian acyclic model (LiNGAM)\cite{shimizu2006linear}, as well as the NOTEARS method\cite{zheng2018dags}. Regarding the data generation, a random graph $G_0$ is firstly generated from two random graph models: Erd$\ddot{o}$s-R$\acute{e}$nyi (ER) graph or scale-free (SF) graph. Then we randomly added extra edges to $G_0$ and got $G_1$ and $G_2$ to simulate the case-control study. In this way, we can build the association between those two graphs: sharing edges are the ones in $G_0$ and the varied edges are those independently added extra edges. Once graphs were generated, we equipped edges with weights according to an uniform distribution and obtained weight matrix $W_i$ for each graph. The generation and alternation of random graphs were realized through the Python package NetworkX. Given $W_i$s, the observation sequences were then sampled on the basis of $X_i = W_i^TX_i + \epsilon_i$ for various distributions of $\epsilon_i$. In this paper, we did this for three types of noise distribution: Gaussian, Exponential, and Gumbel. We repeated above procedure to generate $X_i \in \mathbb{R}^{n\times d}$ in two simulation settings. We generated the first group of simulation data with  $n=300$ and $p=\{10, 20, 50, 100\}$ ; as for the high dimension case, we then generated the second group of simulation data with $n=50$ and $p=\{20, 40, 60, 100\}$.
It has been confirmed in many literatures that scale-free structure is common in many discovered real-world networks such as gene networks, social networks, and the brain functional networks\cite{eguiluz2005scale}\cite{van2008small}.
\subsubsection{Parameters Setting}
For each data set, we ran GES, PC, LiNGAM, NOTEARS and our proposed method, comparing their performance in terms of the reconstruction of the ground truth DAGs for two groups. We employed the following implementations in the experiment:
\begin{itemize}
  \item GES, PC and LinGAM were implemented using the widely used R package \textit{pcalg}, which is available at \url{https://cran.r-project.org/web/packages/pcalg/index.html}.
  \item NOTEARS method was implemented using Python code available at \url{https://github.com/xunzheng/notears}
  \item The proposed method was implemented with Python code, which is available at \url{https://github.com/gmeng92/joint-notears}.
\end{itemize}
There are some further remarks on the compared methods: both PC and GES method output a graph instead of the weight matrix $W$; the graph returned by those two methods is a CPDAG instead of a DAG, which means part of the edges are undirected. During the evaluation, we treated PC and GES method favorably by regarding undirected edges as true positives if there exist directed edges between the same nodes. For the parameter tuning, we followed the same setting according to suggestions in \cite{zheng2018dags}.
\subsubsection{Metrics Used for Comparison}
To evaluate the estimated DAGs (for LiNGAM, NOTEARS and our method) or CPDAGs(for PC and GES method), we used three common metrics: 1) false discovery rate (FDR), 2)
true positive rate (TPR),  and 3) structural Hamming distance (SHD).  The first two measures comes from the confusion matrix, which is defined in Table \ref{Tab:conf_mat}
\begin{table}[htbp]

  \centering
  \caption{Confusion Matrix}
    \begin{tabular}{|cm{5.5em}c|c|}
   \hline
    \multicolumn{2}{|c|}{\multirow{2}[0]{*}{}} & \multicolumn{2}{c|}{True Condition} \\ \cline{3-4}
    \multicolumn{2}{|c|}{} & \multicolumn{1}{b{5em}|}{Condition Postive} & \multicolumn{1}{b{5em}|}{Condition Negative} \\
    \hline
    \multicolumn{1}{|m{5em}}{\multirow{2}[0]{*}{\minitab{Predicted\\ Condition}}} & \multicolumn{1}{|c|}{\minitab[c]{Predicted\\Postive}} & True Postive & False Positive \\
    \cline{2-4}
        & \multicolumn{1}{|c|}{\minitab[c]{Predicted\\Negative}} & False Negative & True Negative \\
          \hline
    \end{tabular}%
 \label{Tab:conf_mat}%
\end{table}%

The SHD is defined as the number of operations needed to convert the estimated graph to the ground truth graph. These operations include edge additions, deletions and direction reversals. Since we consider directed graphs, a distinction between true positives (TP) and reversed edges (R) is needed: the former is estimated with correct direction whereas the latter is not. Likewise, a false positive (FP) is an edge that is not in the undirected skeleton of the true graph. In addition, positive (P) is the set of estimated edges, true (T) is the set of true edges, false (F) is the set of non-edges in the ground truth graph. Finally, let E be the extra edges from the skeleton and M be the missing edges from the skeleton. The four metrics are then given by:
\begin{itemize}
  \item{FDR = (R + FP)/P}
  \item{TPR = TP/T}
  \item{SHD = E + M + R}
\end{itemize}
\subsubsection{Results}
In Fig. \ref{fig:simu_demo}, we gave a demonstration of the joint estimation method on a synthetic data set with two groups and limited sample size($n = p = 20$). By comparing the estimation results with the ground truth DAG, we can clearly see that the joint estimation has improved the estimation accuracy significantly compared to the NOTEARS approach. This demonstrates that structure learning does benefit from the group regularization.

\begin{figure}[htbp]
\centering
\subfloat[$n=300$]{\includegraphics[width=0.45\textwidth]{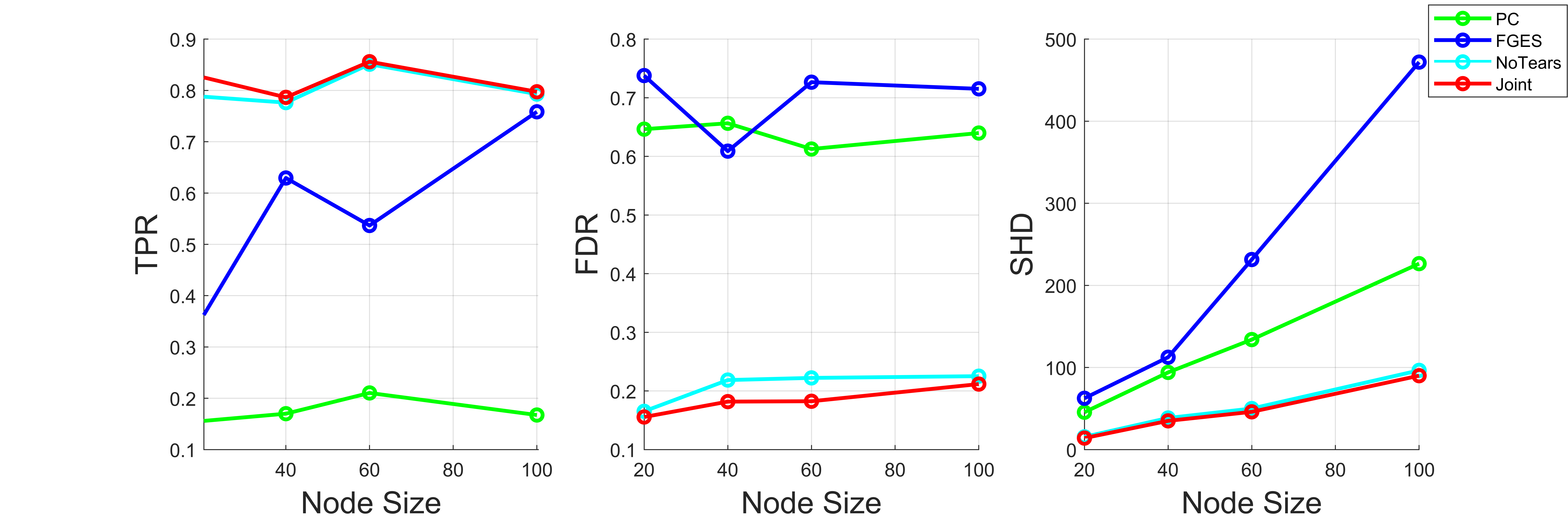}%
\label{fig:simu1}}
\hfil
\subfloat[$n=50$]{\includegraphics[width=0.45\textwidth]{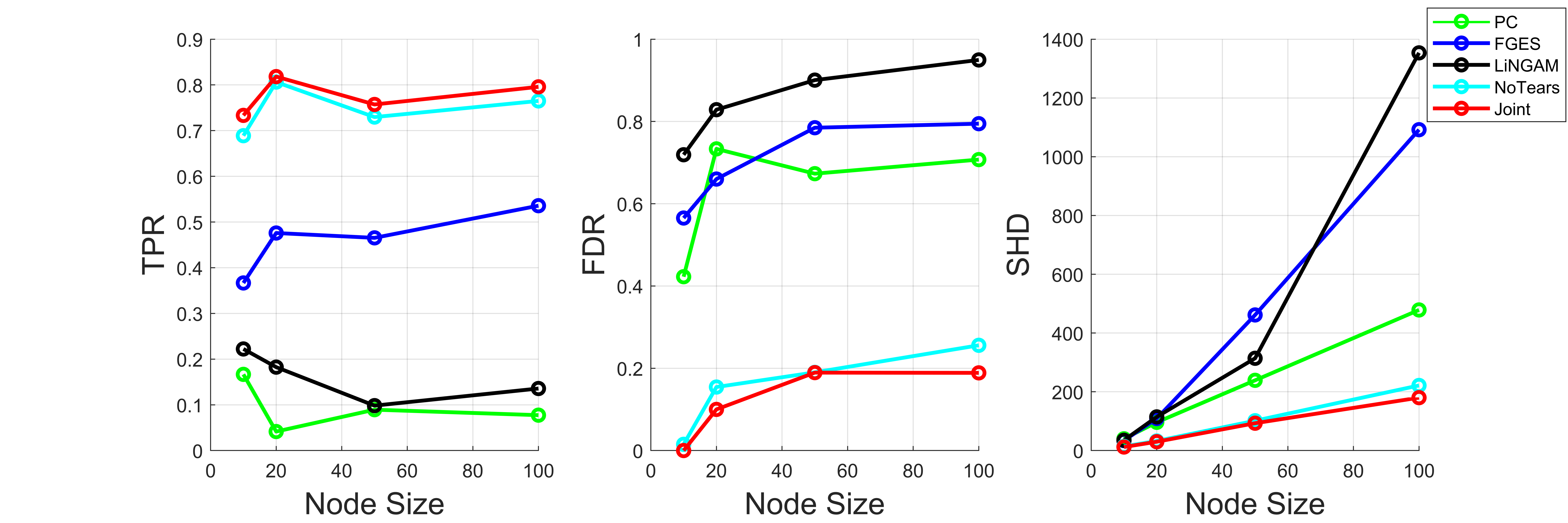}%
\label{fig:simu2}}
\caption{Simulation results for the network. The performance measures, including TPR, FDR and SHD, was reported for two simulation data sets with sample size: (a) $n=300$ and (b) $n=50$.}
\label{fig:simu_main}
\end{figure}

Moreover, we show the comparison results when the noise type is Gaussian, average degree is 4, sample size are $n=300$ and $n=50$ respectively for ER graphs in Fig. \ref{fig:simu_main}. Both NOTEARS method and the proposed joint method have the highest TPR while the lowest FDR and SHD, which shows the superiority of those structure learning framework. It is noticeable that as the sample size decreases and the number of variables grows, the FDR and the SHD increase significantly. But the joint estimation can still efficiently suppress the FDR and SHD to a low level, that is, the estimation accuracy is sustained even in high dimensional scenario. For synthetic data sets with various noise types, graph densities and random graph types, the joint estimation method has outperformed other methods. More details can be found in \textbf{Appendix.C}.
\subsection{On real fMRI Data}
The model was applied to fMRI data in the MIND Clinical Imaging Consortium (MCIC) from Mind Research Network(MRN,\url{www.mrn.org}). The fMRI data were collected from 208 subjects, among them 92 SZ patients (age: 34 $\pm$ 11, 22 females)
and 116 healthy controls (age: 32 $\pm$ 11, 44 females)during a sensory motor task, a block design motor response to auditory stimulation. Specifically, the images were
acquired on a Siemens3T Trio Scanner and 1.5 T Sonata with echo-planar imaging (EPI)
sequences taking parameters (TR $= 2000 ms$, TE $= 30 ms (3.0 T)/40 ms (1.5 T)$, field of
view = $22 cm$, slice thickness = $4 mm$, $1 mm$ skip, 27 slices, acquisition matrix = 64 $\times$ 64, flip angle = $90^{\circ}$). Then the data were pre-processed with SPM5 (\url{http://www.fil.ion.ucl.ac.uk/spm})
and were realigned, spatially normalized and resliced to $3 \times 3 \times 3 mm$, smoothed with a $10 \times 10 \times 10 mm^3$ Gaussian kernel, and analyzed by multiple regression considering the stimulus and their temporal derivatives plus an intercept term as regressors. Finally the stimulus-on versus stimulus-off contrast images were extracted with $53 \times 63 \times 46$ voxels and all the voxels with missing measurements were excluded resulting in 41236 voxels. In order to filter irrelevant information, we further implemented a multiple t-test between case and control groups at the voxel level. Finally, $p = 9816$ voxels were left for analysis and 116 ROIs were extracted based on the Automated Anatomical Labeling brain atlas. For each ROI, we then averaged the beta values of the voxels within that ROI and finally got an $116\times 208$ data matrix.
\subsubsection{Summary of graph theoretical measure}
\begin{figure}[htbp]
\centering
\includegraphics[width=0.5\textwidth]{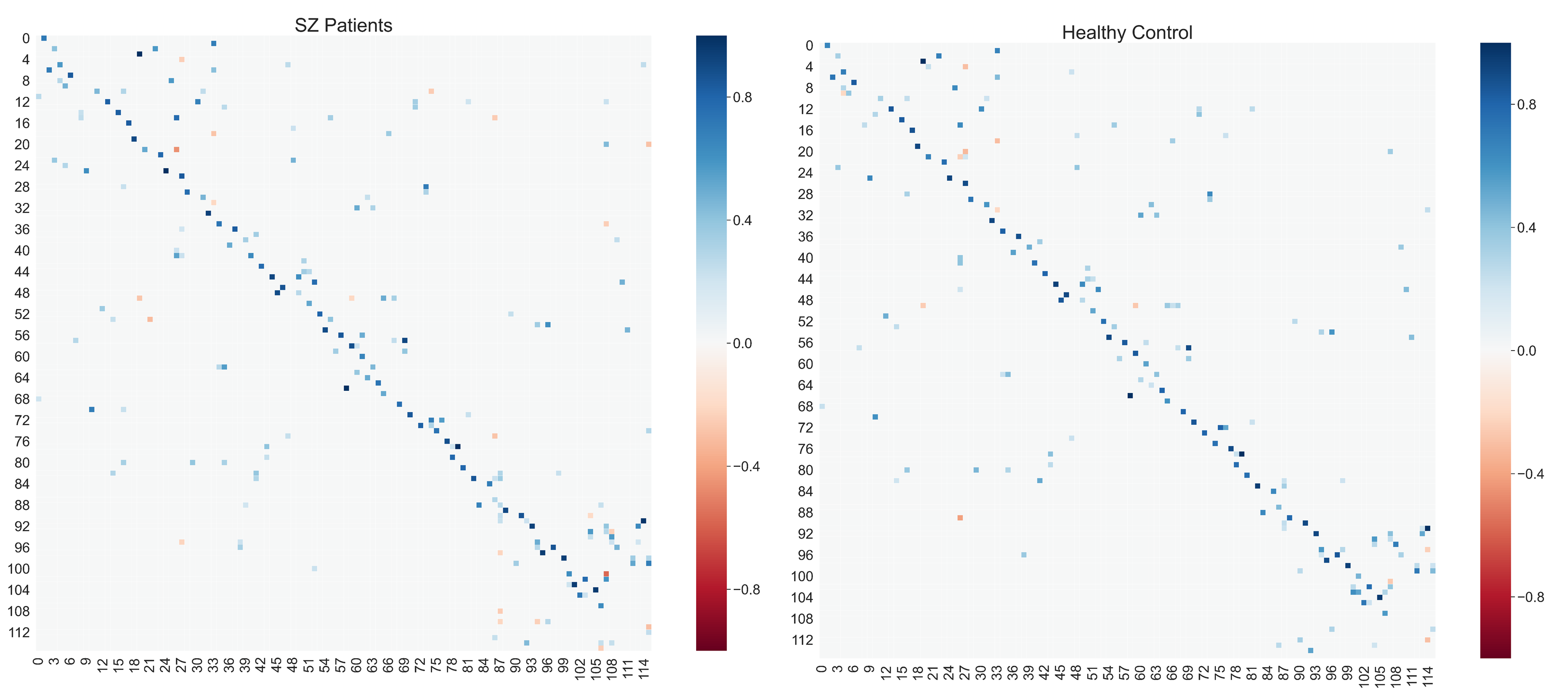}%
\caption{The jointly estimated directed brain networks of SZ patients (right) and healthy controls (left).}
\label{fig:sz_hc}
\end{figure}

After splitting the data into two groups, we fed them into the joint estimation algorithm with $\lambda_1=0.001$ and $\lambda_2=0.01$, and got the corresponding DAGs and the weighted adjacency matrices $\mathbf{W}_1, \mathbf{W}_2$ of SZ patients and healthy controls respectively. The resulted DAGs of two group were visualized in Fig. \ref{fig:sz_hc} with 207 directed edges for SZ group and 187 directed edges for healthy controls. In the first step, we compared the two DAGs using some graph theoretical measures with the purpose of finding the general structural differences between two groups. Since the effect size of the weights still cannot be well estimated, the structures have a higher power and the edge weights are ignored in calculations. Those measures were calculated using the Brain Connectivity Toolbox\cite{rubinov2010complex}. Table \ref{Tab: measures} shows the summary of those measures and the p-values of the permutation test were listed in the last column.
The density is the fraction of present connections to all possible connections and it describe how the dense the graph is. The number of connections to one node is defined as the degree of that node. For a directed graph, those connections include edges coming in and out of the node. In directed network analysis, the in-degree counts the number of edges that come into one individual node, the out-degree counts the number of edges that come out of the node, and the sum-degree counts the total number of edges connected to the node regardless of the direction.
Global efficiency is the measure of functional integration, which reflects the ability to combine specialized information from distributed brain regions. It is defined as the average inverse shortest path length, which represents the ability of the communication between two nodes. The local efficiency is the global efficiency computed on the neighborhood of the node and represents how well the graph communicates locally. Like the transitivity and clustering coefficient, local efficiency is a measure of functional segregation, which reflects the ability for specialized processing to occur within clusters interconnected modules of brain.
Transitivity and clustering coefficient are based on the number of triangles in the graph. The fraction of triangles around an individual node is defined as the clustering coefficient while the ratio of triangles to triplets in the whole graph is defined as the transitivity. The mean clustering coefficient averages the values of all nodes. The assortativity coefficient is a correlation coefficient between
the degrees of all nodes on two opposite ends of a link. Graphs with a positive assortativity coefficient are likely to have a resilient core of mutually interconnected high-degree hubs while a negative value means widely distributed hubs. The mean assortativity coefficient averages the out-degree/indegree,
in-degree/out-degree, out-degree/out-degree and
in-degree/in-degree correlations. The rich club coefficient
at level k is the fraction of edges that connect nodes
of degree k or higher out of the maximum number
of edges that such nodes might share. We used the
maximum of the rich club coefficients among various
levels, which characterizes the tendency for high degree
nodes to be more inter-connected among themselves
than the nodes of a lower degree.
\begin{table}[htbp]
\caption{Comparison of graph theoretical measures between schizophrenia patients (SZ) and healthy control (HC)}
\label{Tab: measures}
\centering
\begin{tabular}{c|c|c|c}
\hline
Group & SZ & HC & p-value\\
\hline
Density & 0.0155 & 0.0139   & <0.01\\
Transitivity & 0.0275 & 0.0250  & 0.03 \\
Mean clustering coefficient & 0.0308 & 0.0290 & <0.01\\
Maximum rich club coefficient & 0.9290 & 0.9274  & 0.54\\
Global efficiency & 0.0314 & 0.0351  & <0.01\\
Mean local efficiency & 0.0419 & 0.0431  & <0.01\\
Assortativity coefficient &-0.0158&  -0.0668&  0.27\\
\hline
\end{tabular}
\end{table}
\subsubsection{Hub nodes identification}
To get better understanding of the estimated directed brain networks from two groups, we further identified the hub nodes as the nodes with significantly higher degrees than average. More specifically, we defined the nodes whose degree is higher than mean degree by three standard deviations. According to previous context, we have three type of degrees here: in-degree, out-degree and sum-degree. As a result, we can similarly define three types of hub nodes for directed networks: in-hub, out-hub and sum-hub.  Since some nodes were identified as sum-hubs because it has a high in-degree or out-degree, we only keep the sum-hubs if they are not in-hubs or out-hubs.

In the healthy control group, we were able to identify two in-hubs located in gyrus rectus(REC.L) and cerebellum(CRBL10.R), one out-hub in superior occipital lobe(SOG.R) and one sum-hub in cerebellum(CRBL45.L). In the SZ patient group, we identified three in-hubs located in right temporal pole(TPOmid.R), cerebellum(CRBL10.R) and vermis(Vermis10) as well as two out-hubs in inferior frontal gyrus(IFGtriang.L) and cerebellum(CRBL3.R). Those results were summarized in Table \ref{tab:hubs}.
\begin{table}[htbp]
  \centering
  \caption{Identified hub nodes for case and control group}
  \resizebox{0.48\textwidth}{!}{
    \begin{tabular}{l|c|l|c|c}
    \hline
    \multicolumn{5}{c}{SZ Patients} \\ \hline
          & Index & ROI   & MNI coordinate & \multicolumn{1}{l}{Degree} \\ \hline
    \multirow{3}[0]{*}{In-hub} & 88    & TPOmid.R  & (44.22, 14.55, -32.23) & 8 \\
          & 108   & CRBL10.R  & (25.99, -33.84, -41.35) & 7 \\
          & 116   & Vermis10  & (0.36, -45.8, -31.68) & 6 \\ \hline
    \multirow{2}[0]{*}{Out-hub} & 13    & IFGtriang.L  & (-45.58, 29.91, 13.99) & 5 \\
          & 96    & CRBL3.R  & (12.32, -34.47, -19.39) & 5 \\ \hline
    \multicolumn{5}{c}{Healthy Control} \\ \hline
          & Index & ROI   & MNI coordinate & \multicolumn{1}{l}{Degree} \\ \hline
    \multirow{2}[0]{*}{In-hub} & 27    & REC.L  & (-5.08, 37.07, -18.14) & 6 \\
          & 108   & CRBL10.R  & (25.99, -33.84, -41.35) & 5 \\ \hline
    \multicolumn{1}{l|}{Out-hub} & 50    & SOG.R  & (24.29, -80.85, 30.59) & 5 \\ \hline
    \multicolumn{1}{l|}{Sum-Hub} & 97    & CRBL45.L  & (-15, -43.49, -16.93) & 6 \\ \hline
    \end{tabular}}%
  \label{tab:hubs}%
\end{table}%

\subsubsection{Variations of edges between groups}
\begin{figure}[htbp]
\centering
\includegraphics[width=0.48\textwidth]{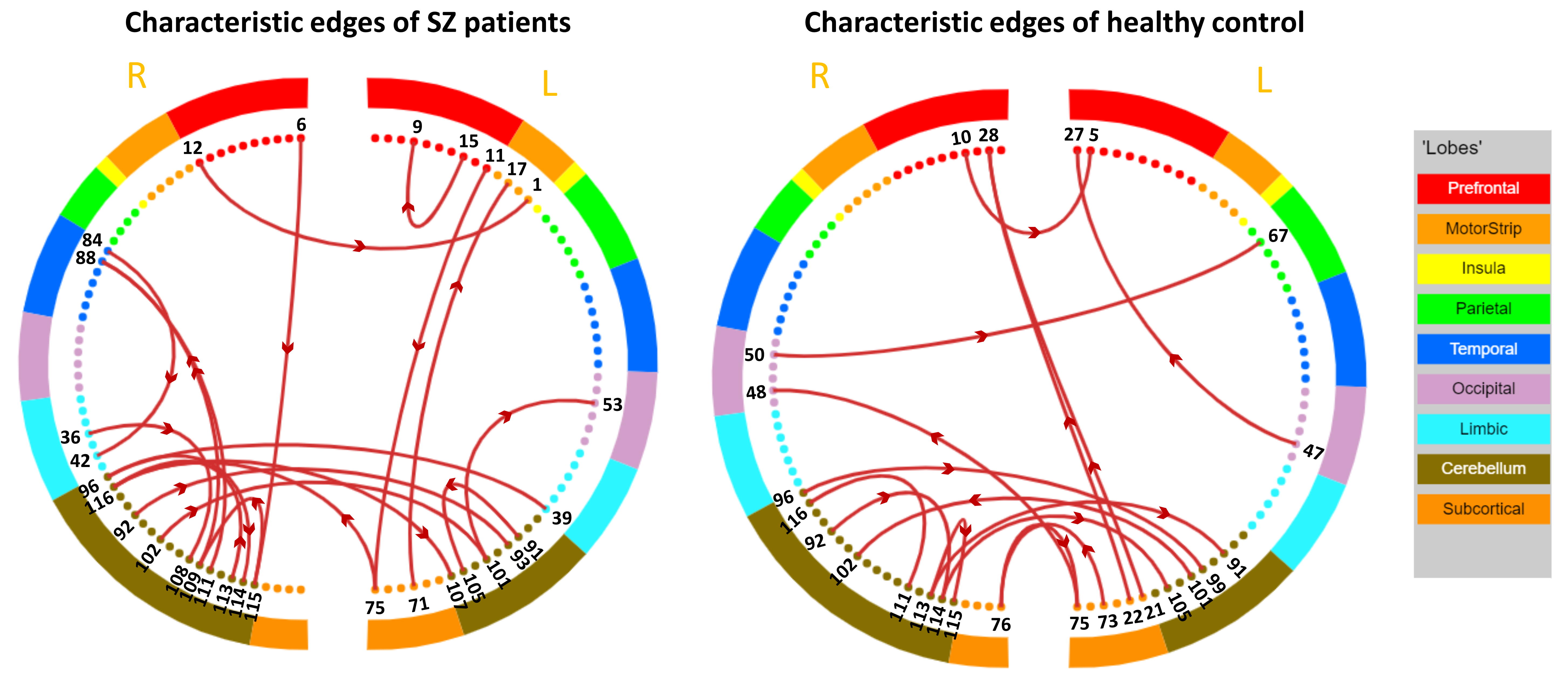}%
\caption{The Characteristic edges in directed networks of SZ patients and healthy controls. Arrows represents the direction of edges and numbers represent the ROI indexes which can be found in \textbf{Appendix 4.3} Table 1}
\label{fig:ex_edges}
\end{figure}

The graph measures only give a brief summary of the variation between two groups. In this study, we are more interested in those particular edges or nodes shared between two groups or characterized themselves. Hence, we further did a permutation test for the comparison. 100 permutation tests were performed and the subjects were randomly shuffled into two groups - one with 92 subjects and another with 116 subjects. Then we extracted the significantly different edges (p < 0.01 with FDR corrected value at q = 0.05) in case and control groups. For each group, the edges that exclusively exist within one group were marked as the characteristic edges(CtEs). Finally, 19 and 15 edges were identified as CtEs in SZ patients and healthy controls respectively, which are illustrated in Figure \ref{fig:ex_edges}.

\subsubsection{Comparison with undirected graphs}
\begin{figure}[htbp]
\centering
\includegraphics[width=0.48\textwidth]{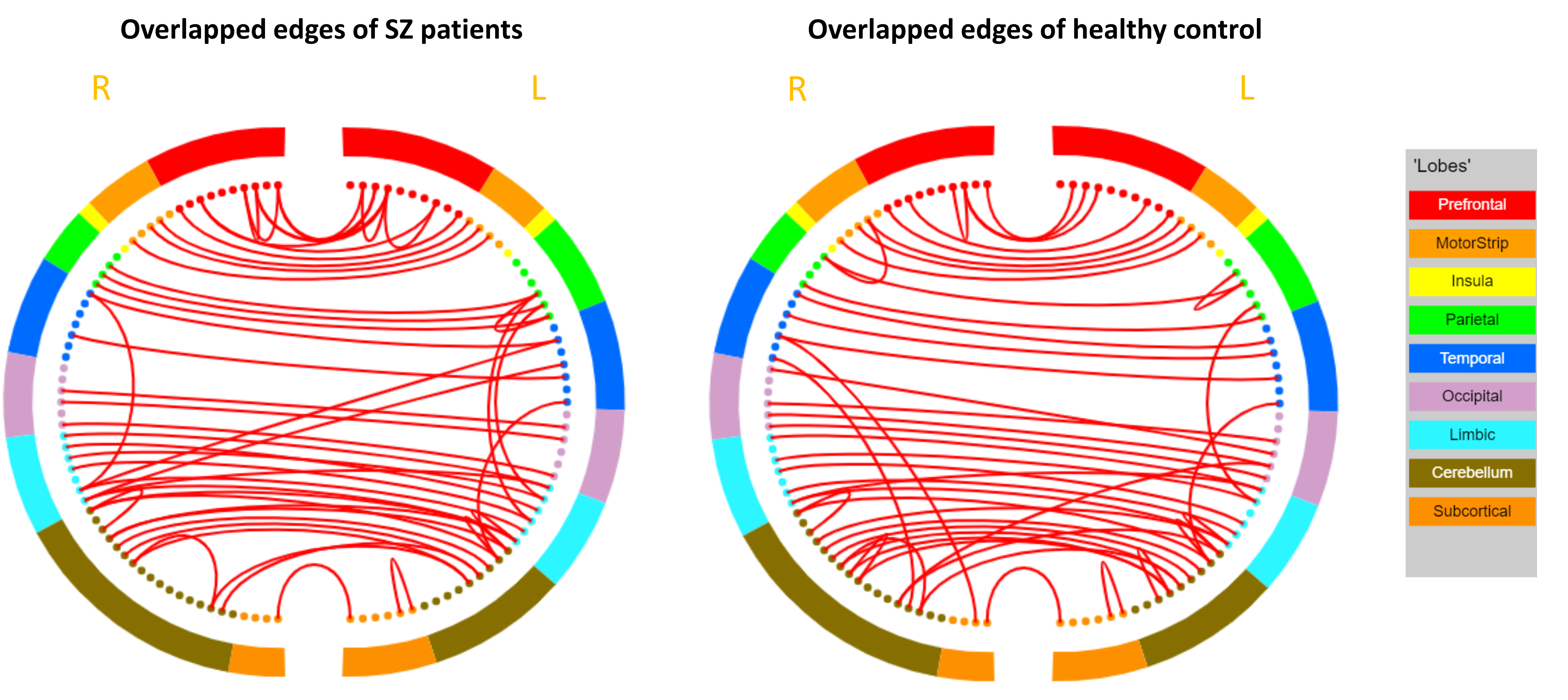}%
\caption{The overlapped edges between directed networks and undirected networks of SZ patients and healthy controls.}
\label{fig:overlap_edge}
\end{figure}

\begin{figure}[htbp]
\centering
\includegraphics[width=0.48\textwidth]{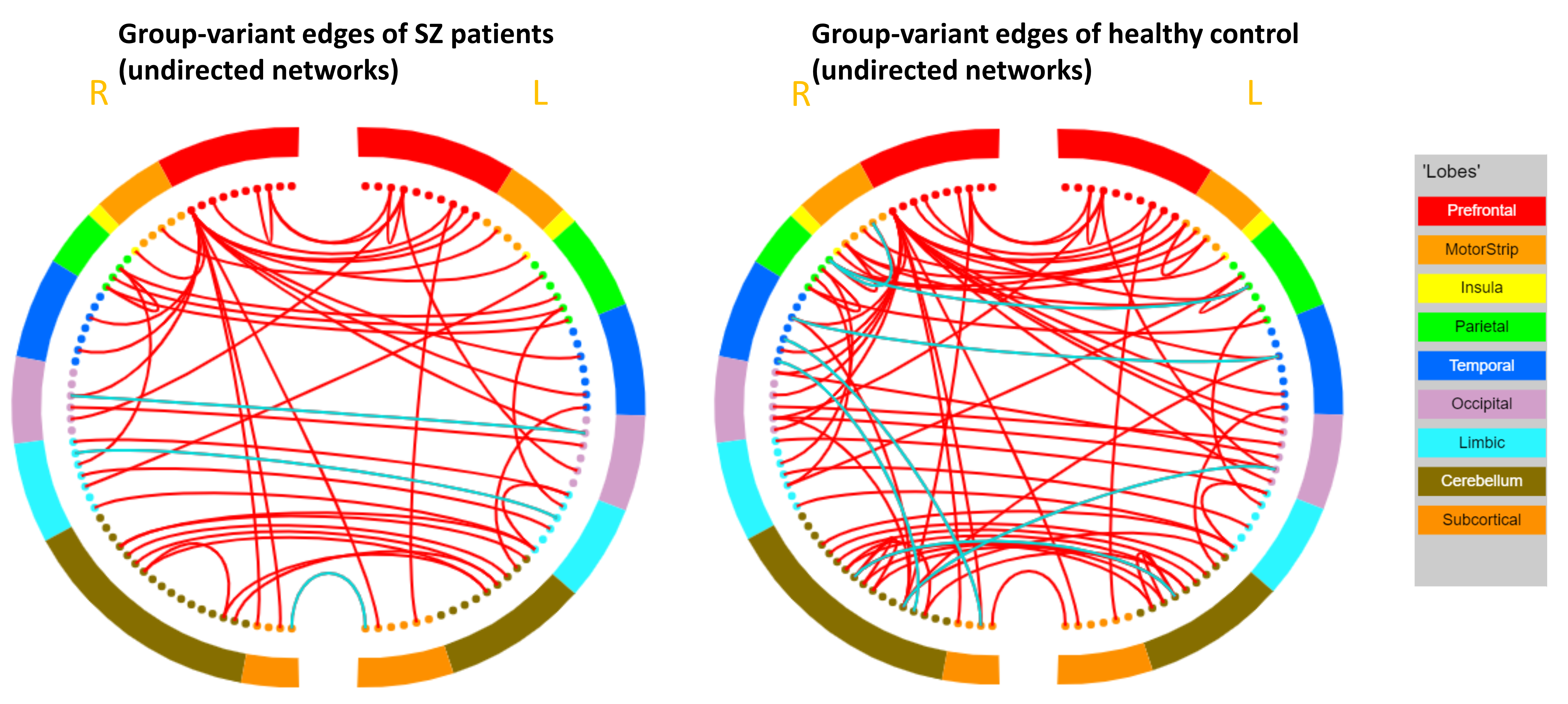}%
\caption{The group-variant edges between SZ patients and healthy controls in undirected networks. Lines colored with blue indicate their appearance in respective directed networks.}
\label{fig:groupvar_edge}
\end{figure}

Another interesting question one may ask is the added values of the directed graphs we obtained over the undirected graphs measured with Pearson correlation or partial correlation, which are often named as functional connectivity (FC). Since edges from both graphical models represent conditional dependency in spite of different angles, there are considerable overlaps between two graphs. Recent progress also suggests that the FC can be utilized as a prior to benefit the causal inference\cite{reid2019advancing}. Specially, we used the $psi-$learning method to estimate the undirected brain networks for SZ patients and healthy controls separately. $\psi-$ learning is a novel method for high-dimensional Gaussian Graphical Model (GGM)\cite{liang2015equivalent}. Compared to other partially correlation based methods, it can help ease computational burden and provide more accurate inference for the underlying networks. This method has also been proven to be an equivalent measure of the partial correlation coefficient and thus is flexible for network comparison through statistical tests. In this paper, we used the same implementation as in \cite{zhang2018aberrant} with R package "\textit{equSA}". The parameters were set as $\alpha_1 = 0.1$ and $\alpha_2  = 0.01$ as suggested in \cite{liang2015equivalent, zhang2018aberrant}. For the group comparison, we set the significance level at 0.01 and further corrected the FDR with $q=0.05$.
\begin{table}[htbp]
  \centering
  \caption{Identified hub nodes for SZ patients and healthy control using undirected networks estimation}
  \resizebox{0.48\textwidth}{!}{
    \begin{tabular}{c|l|c|c}
    \hline
    \multicolumn{4}{c}{SZ Patients} \\\hline
    Index & ROI   & MNI coordinate & \multicolumn{1}{l}{Degree} \\\hline
    12    & IFGoperc.R & (-50.2, 14.98, 21.41) & 19 \\
    39    & PHG.L & (-21.17, -15.95, -20.7) & 15 \\
    40    & PHG.R & (25.38,-15.15,-20.47) & 21 \\
    \hline
    \multicolumn{4}{c}{Healthy Control } \\\hline
    Index & ROI   & MNI coordinate & \multicolumn{1}{l}{Degree} \\\hline
    12    & IFGoperc.R & (-50.2, 14.98, 21.41) & 21 \\
    111   & Vermis45 & (1.22, --52.36, -6.11) & 7 \\\hline
    \end{tabular}}%
  \label{tab:hubs_und}%
\end{table}%
Finally, we got 148 edges and 104 edges in SZ patients and healthy controls using $\psi-$learning method respectively. To compare the undirected networks with the directed networks, we mainly checked three aspects: the hub nodes in undirected networks, the overlapped edges and the detected group variances between undirected networks and directed networks. The identified hub nodes is listed in Table \ref{tab:hubs_und}. Opercular region of inferior frontal gyrus(IFGoperc.R) at the right lobe was identified as hub nodes in both groups. In SZ patients, parahippocampus on both lobes (PHG.L and PHG.R) were identified as hub nodes. In healthy control, a region at vermis (VERMIS45) was also identified as hub node. It should be noticed that there is no overlapped hub nodes between directed networks and undirected networks. The overlapped edges between directed networks and undirected networks for both groups were visualized in Figure \ref{fig:overlap_edge}. There is no CtEs appearance in either group. For the pairwise comparison, significantly different edges between two groups were visualized in Figure \ref{fig:groupvar_edge}. As a comparison, we also highlighted the overlapped edges from both directed networks and undirected networks with blue color in that figure. However, those connections do not occupy considerable percentage shared in both groups ($12.96\%$ in SZ patients and $15.57\%$ in healthy controls).

\section{Discussion}\label{sec:4}
\subsection{Significance of results in MCIC study}
The application of functional connectivity to study schizophrenia
appears to be a very powerful approach since it provides a way to
study aberrant connectivity between sets of regions (networks),
which is thought to be a core feature of schizophrenia. However, the directed functional connectivity is rarely used as biomarkers in the schizophrenia study. From the comparison between two groups, we found that the directed brain network is significantly denser in schizophrenia patients. Moreover, the mean clustering coefficient of schizophrenia patients is smaller than that of the healthy controls. With a denser network, we observed that both the global and local efficiency of schizophrenia patients are significantly smaller than those of healthy controls. The functional integration represents the capability to combine information from distributed brain regions. This demonstrates the directed brain network of schizophrenia patients is less efficient in terms of brain region communication either globally or locally in the auditory task.

Among the hub nodes identified in two groups, we also observed disruptions of the hub structures in schizophrenia patients. The gyrus retus (REC.L), as an in-hub on the left hemisphere, is missing in SZ patients. Previous study\cite{zhao2018structural} has demonstrated that pronounced gray matter volume decline at gyrus rectus has been observed in SZ patients, which supports our findings with a structural foundation. Another study\cite{li2019dysconnectivity} also showed that the reduced functional connectivity has been observed between gyrus rectus and other regions within self-referential processing network. Inferior frontal gyrus, triangular at left hemisphere (IFGtriang.L), has been shown to have increased activity during auditory hallucinations\cite{mcguire1993increased}, and the Theory of Mind(ToM) deficits \cite{das2012mentalizing} in schizophrenia patients. Our result further implies that the hyperconnectivity related to this region is potentially an important biomarker of schizophrenia. The behavior of temporal pole on the right hemisphere (TPOmid.R) is also distinct between HC and SZP, which agrees with the finding in \cite{gao2020functional}, \cite{chen2019cortico}.

In addition to the abnormal hub structures in SZ patients, we also have several findings in coordination with previous studies by comparing the CtEs extracted from both groups. For example, previous research\cite{kiparizoska2017disrupted} revealed that olfactory bulbs disconnectivity of olfactory regions in schizophrenia may account for olfactory dysfunction and disrupted integration with other sensory modalities in SZ patients. The missing connections to the olfactory cortex in CtEs of SZ patients also confirm this findings. In the subcortical regions, pallidum, putamen and caudate work together to communicate with the subthalamic nucleus\cite{alexander1990functional}. Our findings illustrated the missing communications within and between putamen and pallidum, which is in line with the findings in \cite{wang2017su101},\cite{cui2016putamen}.

From Table \ref{tab:hubs_und}, parahippocampus on both sides are identified as hub nodes in undirected functional connectivity for SZ patients. The parahippocampus plays an important role in the encoding and recognition of environmental scenes. Previous research\cite{du2018dynamic} have discovered the abnormal connections between parahippocampus and temporal pole in early stage SZ patients. Several other researches also revealed the aberrant increased connectivity between parahippocampus and other limbic areas\cite{hua2020disrupted}, including hippocampus\cite{kraguljac2016aberrant}.

\begin{figure}
  \centering
  \includegraphics[width=0.30\textwidth]{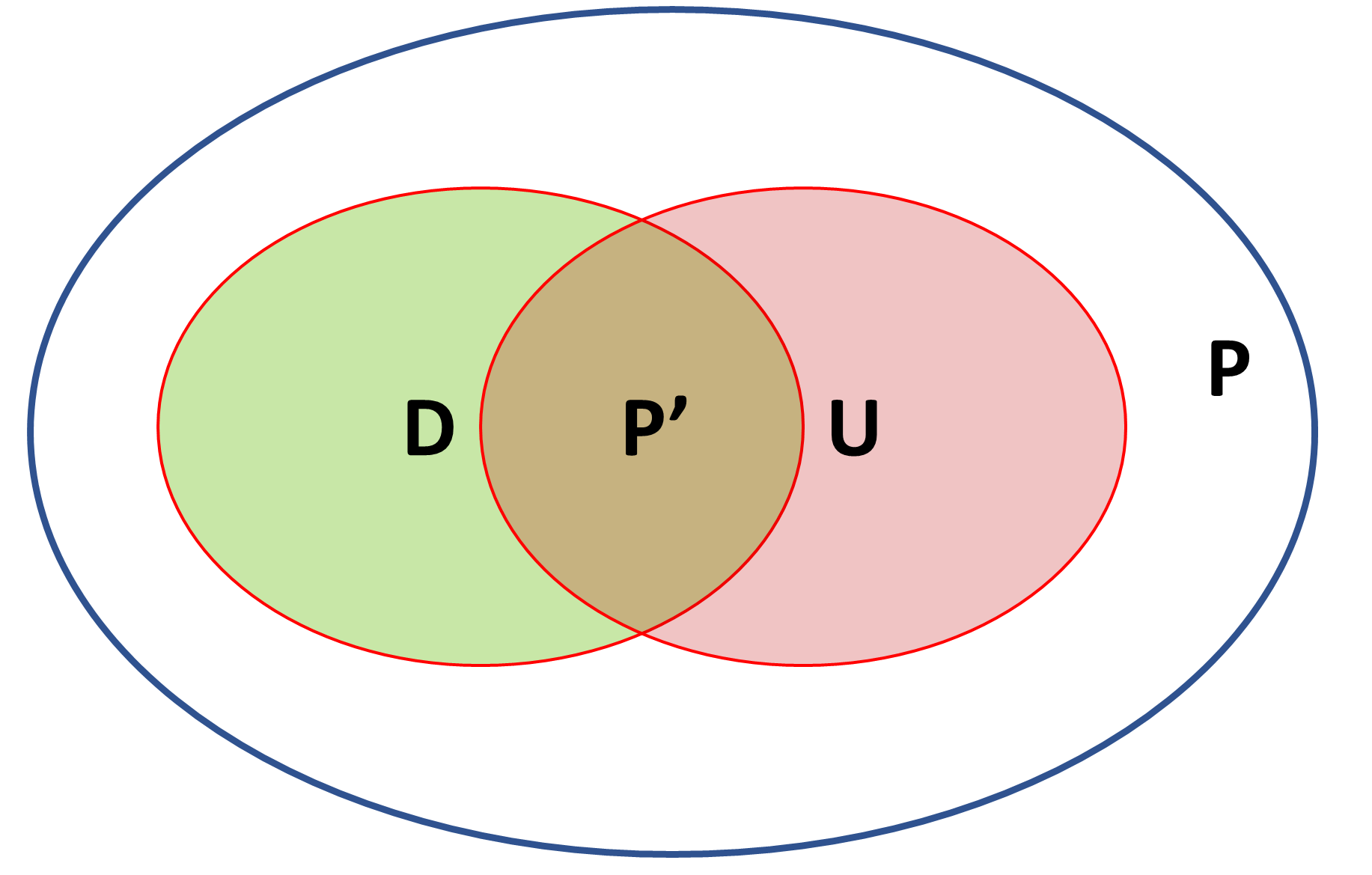}\\
  \caption{Venn diagram illustrating the set of all distributions
P over a given set of variables, together with the
set of distributions D that can be represented as a
perfect map using a directed graph, and the set U
that can be represented as a perfect map using an
undirected graph.}\label{fig:venn_diag}
\end{figure}

When we compared the finding of directed graphical model and undirected graphical model, we found that the overlapped brain networks indicated by both methods cannot represent the feature of whole brain networks, especially the connection variations between SZ patients and healthy controls. This implies that the features from undirected networks and directed networks may not be regarded as the similar features in depicting the interactions between brain regions. In fact, the underlying distributions of two graphical modelling framework are intrinsically different. Figure \ref{fig:venn_diag} demonstrates the relationship between two models. Only the distributions falling in the intersection of directed graphs (D) and undirected graphs (U) are perfectly mapped through both graphical models. It is clearly that we do not know the underlying distribution of brain activities. We speculate that either method can only provide a sketch of the ground truth from one particular perspective. Hence, we reported the pairwise comparisons results of both methods.
\subsection{Limitations}
 Currently, both the NOTEARS and the proposed method were optimized with second order approximation with a global searching over the feasible space, which is time consuming if we want to extend those methods to application with hundreds of variables, a typical case in brain imaging study. For example, it is impossible to directly apply the method for voxel-wise network analysis. It will be desirable if we can find a way to reduce the search space in the model. In fact, the model itself has a flexible setting to incorporate the prior information of the network in the optimization, which will significantly alleviate the computation burden. The prior information includes some known connections/disconnections and the directions of connections. Another limitation lies in the variances of the results that we found between the directed network method and the undirected network method. At this stage, most of whole brain functional connectivity studies are based on either directed or undirected network analysis. It will be meaningful to cross-validate the findings with more data sets.
\section{Conclusion}\label{sec:5}
In this paper, we proposes a joint DAGs model for structural learning via a continuous optimization framework. By encouraging the group similarity in the DAG structure, the model can be used to find DAGs for multiple groups simultaneously. The efficiency of the proposed model was tested using simulation data sets with a wide range of varieties including density levels, noise types, etc. By comparing with other structural learning methods, the proposed model have high TPR and lower FDR, especially in high dimensional cases. As a demonstration of its application, the proposed model was used to detect the abnormal network structures in schizophrenia patients in task fMRI data from MCIC. The lower global and local efficiency demonstrated the deficiency of functional integration in SZ patients. In addition, we also identified disrupted hub structures and characteristic edges in SZ patients. Several of the findings have been in line with previous schizophrenia studies. Finally, we compared the results of directed networks to those with the undirected networks modelling method. Although both network modelling methods can differentiate the key alterations between schizophrenia patients and healthy controls, the extracted networks' feature provide different perspectives and biological significance.
\section*{Acknowledgment}
\par The authors would like to thank the partial support by NIH (P20GM109068, R01MH104680, R01MH107354, R01MH103220, R01EB020407) and NSF (\#1539067).
\ifCLASSOPTIONcaptionsoff
  \newpage
\fi

\bibliographystyle{IEEEtran}
\bibliography{joint_notear_ref}
\end{document}